\documentclass[10pt,twocolumn,letterpaper]{article}

\usepackage{cvpr}              

\usepackage[dvipsnames]{xcolor}

\usepackage{algorithm}
\usepackage{algorithmic}
\usepackage{amsmath}
\usepackage{amssymb}
\usepackage{mathtools}
\usepackage{booktabs}
\usepackage{tabularx}
\usepackage[export]{adjustbox}
\usepackage{multirow}
\usepackage{colortbl}
\usepackage{bbding}
\usepackage{amssymb}
\usepackage{pifont}

\definecolor{mygray}{gray}{.9}

\definecolor{cvprblue}{rgb}{0.21,0.49,0.74}
\usepackage[pagebackref,breaklinks,colorlinks,citecolor=cvprblue]{hyperref}

\usepackage{makecell}
\usepackage[accsupp]{axessibility}

\definecolor{mygray}{gray}{.9}
\definecolor{myyellow}{rgb}{0.71, 0.55, 0.0}
\definecolor{myblue}{rgb}{0.0, 0.71, 0.71}
\definecolor{color2}{rgb}{0.55, 0.71, 0.0}
\definecolor{color1}{rgb}{0.98, 0.81, 0.69}
\definecolor{color3}{rgb}{1.0, 0.6, 0.4}
\definecolor{color4}{rgb}{0.29, 0.59, 0.82}
\definecolor{myred}{rgb}{0.81, 0.0, 0.0}
\definecolor{myblue}{rgb}{0.0, 0.71, 0.71}
\definecolor{mygreen}{rgb}{0.0, 0.51, 0.0}

\newcommand{\encoder}{\mathcal{E}}
\newcommand{\decoder}{\mathcal{D}}

\def\1{\bm{1}}

\def\paperID{3947} 
\def\confName{CVPR}
\def\confYear{2024}

\title{Self-Adaptive Reality-Guided Diffusion for Artifact-Free Super-Resolution}

\author{
\vspace{2pt}
Qingping Zheng\textsuperscript{1}, 
Ling Zheng\textsuperscript{3}, 
Yuanfan Guo\textsuperscript{2},
Ying Li\textsuperscript{1}\thanks{Corresponding author}\ ,
Songcen Xu\textsuperscript{2}, 
Jiankang Deng\textsuperscript{2}, 
Hang Xu\textsuperscript{2}
\\
\small \textsuperscript{1}Northwestern Polytechnical University
\quad
\textsuperscript{2}Huawei Noah’s Ark Lab
\quad 
\textsuperscript{3}Tsinghua-Fuzhou Institute for Data Technology
\\
\vspace{-5mm}
}

\begin{document}
\maketitle
\begin{abstract}
Artifact-free super-resolution (SR) aims to translate low-resolution images into their high-resolution counterparts with a strict integrity of the original content, eliminating any distortions or synthetic details.
While traditional diffusion-based SR techniques have demonstrated remarkable abilities to enhance image detail, they are prone to artifact introduction during iterative procedures. 
Such artifacts, ranging from trivial noise to unauthentic textures, deviate from the true structure of the source image, thus challenging the integrity of the super-resolution process.
In this work, we propose \textbf{S}elf-\textbf{A}daptive \textbf{R}eality-\textbf{G}uided \textbf{D}iffusion (SARGD), a training-free method that delves into the latent space to effectively identify and mitigate the propagation of artifacts.
Our SARGD begins by using an artifact detector to identify implausible pixels, creating a binary mask that highlights artifacts. 
Following this, the Reality Guidance Refinement (RGR) process refines artifacts by integrating this mask with realistic latent representations, improving alignment with the original image. Nonetheless, initial realistic-latent representations from lower-quality images result in over-smoothing in the final output. To address this, we introduce a Self-Adaptive Guidance (SAG) mechanism. It dynamically computes a reality score, enhancing the sharpness of the realistic latent. These alternating mechanisms collectively achieve artifact-free super-resolution.
Extensive experiments demonstrate the superiority of our method, delivering detailed artifact-free high-resolution images while reducing sampling steps by 2$\times$. 
We release our code at
\url{https://github.com/ProAirVerse/Self-Adaptive-Guidance-Diffusion.git}.
\end{abstract}    
\begin{figure}
    \centering
    \includegraphics[width=\linewidth]{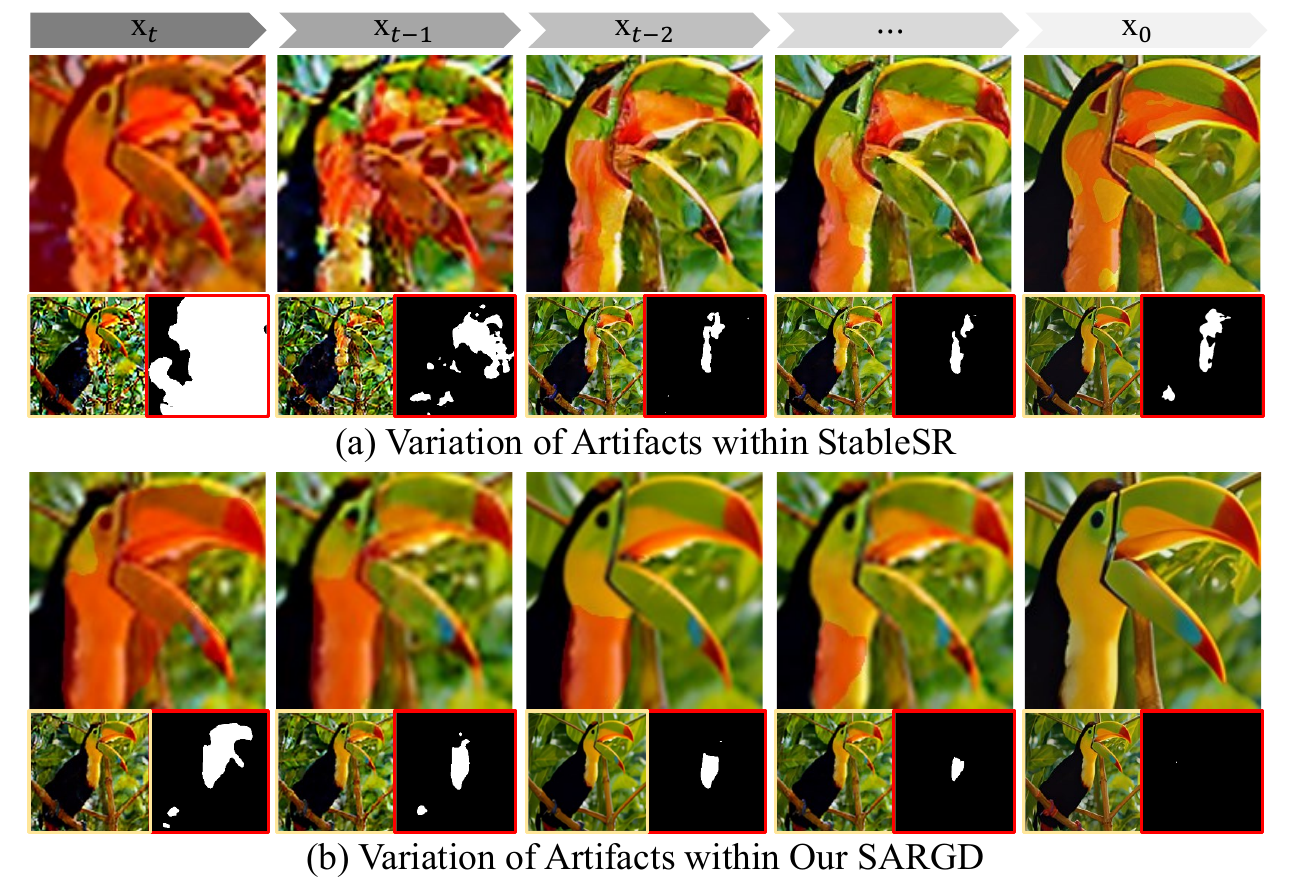}
    \vspace{-8.0mm}
    \caption{\textbf{Visual comparison of artifact variability between (a) StableSR and (b) Our SARGD method.} 
    In the lower panel, the left image is the decoded output, while the right image is the artifact mask.
    Regions containing artifacts are highlighted in {\color{myred} red}, utilizing a binary map to serve as the artifact mask. 
    Our SARGD method exhibits a superior capability in reducing artifacts.}
    \label{fig:intro}
    \vspace{-5.0mm}
\end{figure}
\section{Introduction}
\label{sec:intro}

Super-resolution (SR) techniques strive to restore high-resolution images from their corresponding low-resolution counterparts~\cite{shi2016real,dong2015image}.
This process is inherently ill-posed since the same LR image can yield many plausible HR solutions.
Traditional deep learning SR methods, leveraging techniques like squeeze layers~\cite{zhang2021mr}, transposed convolutions~\cite{yang2018drfn}, and sub-pixel convolutions~\cite{shi2016real}, bind themselves to fixed upsampling scales, limiting output resolution adjustments without architectural changes. 
This causes inflexibility in real-world applications.
To tackle the inherent one-to-many inverse challenge in SR, it is intuitive to envision the solution space as a distribution. Yet, this perspective often prompts generative models to introduce details that deviate from the authentic content of the original low-resolution image.
As a result, a wide range of generative-based SR approaches have emerged to address this intrinsically ill-posed problem,  particularly in diffusion models.

Diffusion-based SR (Diffusion-SR for short) has exhibited significant potential in improving the resolution of images from their low-resolution counterparts. 
SRDiff~\cite{li2022srdiff} is the first diffusion-based model for image super-resolution, progressively transforming the Gaussian noise into a high-resolution (HR) image conditioned on its low-resolution (LR) input through a Markov chain. 
Building on the foundational concept of denoising score matching~\cite{song2019generative}, Saharia~\etal~\cite{saharia2022image} propose SR3 model.  This model advances the field of super-resolution by learning an empirical data distribution and leveraging U-Net architecture~\cite{ronneberger2015u} to progressively denoise and refine the image output.
However, these methods fail to preserve accurate pixel-wise image structures, often necessitating further training or additional skipped connections for detail reproduction.
While the pixel-aware stable diffusion network proposed by Yang~\etal~\cite{yang2023pixel} significantly improves the robustness of the super-resolution process, it concurrently introduces the challenge of generating implausible content.
In contrast, StableSR~\cite{wang2023exploiting}, the current state-of-the-art, leverages pre-trained diffusion priors to retain image details, avoiding assumptions about image degradation.

Nevertheless, as depicted in Figure~\ref{fig:intro}, contemporary diffusion-based StableSR models invariably generate implausible content, frequently resulting in perceptual artifacts within specific areas of the image.
Indeed, these artifacts frequently stem from the fundamental use of Denoising Diffusion Probabilistic Models (DDPMs)~~\cite{ho2020denoising,sohl2015deep} in modern diffusion models.
Specifically, DDPMs, categorized as Partial Differential Equation (PDE)-based models~\cite{chung2023diffusion}, are designed to diffuse pixel intensities in order to reduce noise levels.
They can enhance the preservation of edges and textural features by integrating non-linear terms, thus improving image details. 
However, despite these advancements, the introduction of non-linearity can occasionally yield unexpected outcomes, particularly in areas with intricate textures or noise patterns.
This unintended consequence can result in a loss of fine details and a degradation of essential image elements such as edges and textures, especially when the diffusion process is overly aggressive or inadequately constrained. Additionally, any deficiencies in the model's architecture or parameter settings can be amplified during diffusion, aggravating artifact problems and producing results that deviate from the desired photorealistic quality~\cite {Wang_2023_CVPR}.
These challenges pose a significant obstacle to the preservation of the authentic detail and clarity expected in high-resolution images.

In this work, we propose \textit{\textbf{S}elf-\textbf{A}daptive \textbf{R}eality-\textbf{G}uided \textbf{D}iffusion} (SARGD), a pioneering training-free approach, aiming to produce ``artifact-free" SR images.
In particular, SARGD operates through two stages: Reality-Guided Refinement (RGR) to reduce artifacts and Self-Adaptive Guidance (SAG) to boost image fidelity.
During the initial RGR phase, an artifact detector is employed to identify unrealistic pixels within the latent space at each inference step, leading to a binary artifact mask. 
This mask collaborates with a realistic latent guidance extracted from an upscaled LR image to refine artifacts and enhance intrinsic details.
However, depending on upscaled LR inputs for guidance can potentially result in over-smoothed images, as it may overlook high-frequency details that are essential for SR.
To alleviate this limitation, the subsequent SAG phase introduces a `reality score to self-adjust the authenticity of the latent guidance, significantly boosting the accuracy of detail and texture representation.
Two phases alternate to progressively achieve high-fidelity super-resolution, effectively eliminating artifacts in the diffusion process.
To sum up, the contributions of this paper are as follows:
\begin{itemize}
    \item We introduce a Reality-Guided Refinement (RGR) strategy, designed to identify and rectify artifacts during the diffusion process with the aid of a reality guidance mechanism, effectively eliminating such artifacts.
    
    \item We propose a Self-Adaptive Guidance (SAG) mechanism that iteratively enhances the realistic-latent guidance, addressing over-smoothing issues and improving the fidelity and authenticity of the reference latent guidance.

    \item We are the first to develop the \emph{\textbf{S}elf-\textbf{A}daptive \textbf{R}eality-\textbf{G}uided \textbf{D}iffusion} (SARGD), a training-free approach that sets a new benchmark by effectively dealing with artifacts and over-smoothing issues. Additionally, it outperforms StableSR in image quality while reducing the inference time by 2$\times$ for super-resolution tasks.
\end{itemize}
\begin{figure*}
    \centering
    \includegraphics[width=\textwidth]{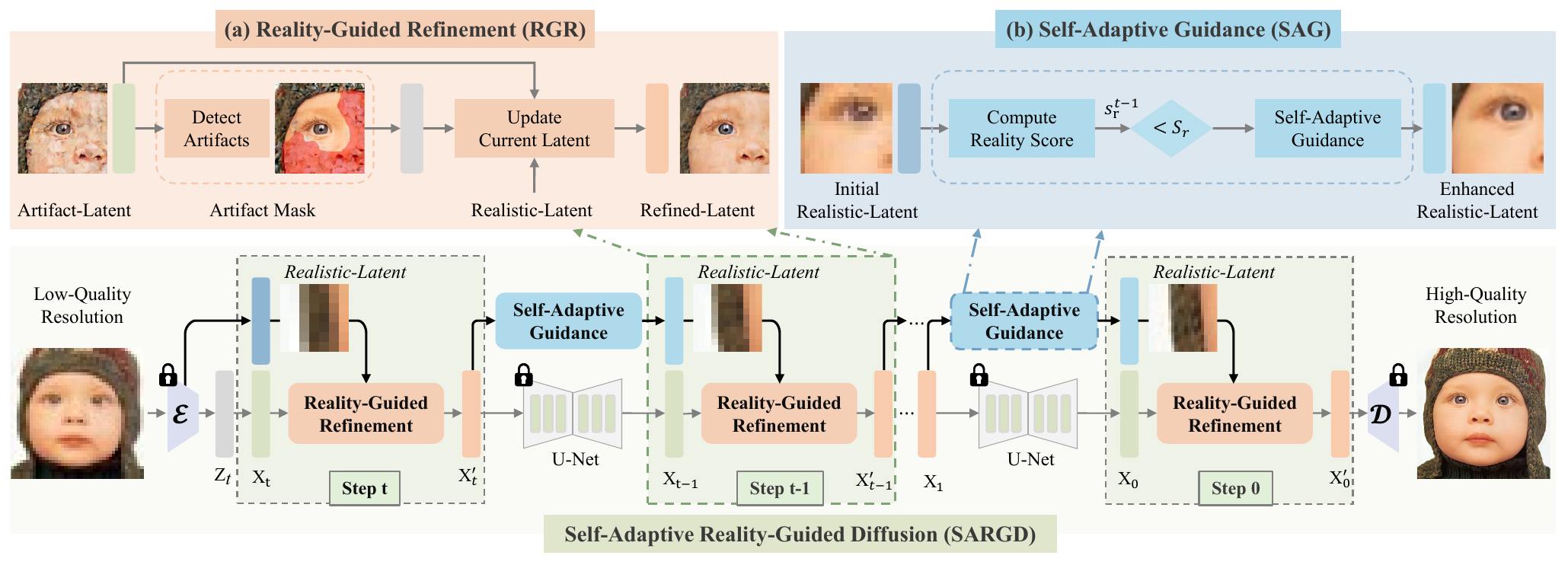}
    \vspace{-8.0mm}
    \caption{\textbf{Self-Adaptive Reality-Guided Diffusion (SARGD)} for artifact-free super-resolution. Our proposed SARGD is a training-free approach that consists of two principal components: 1) a Reality-Guided Refinement (\textbf{RGR}) that identifies and corrects artifacts within the latent representation by using a realistic latent as a guide to maintain the inherent details of the original image during the diffusion process, and 2) a Self-Adaptive Guidance (\textbf{SAG}) mechanism that enhances the fidelity of the initial realistic latent guidance, derived from upscaled low-resolution images, thereby effectively addressing the issue of over-smoothing in the final outputs.}
    \label{fig:framework}
    \vspace{-3.0mm}
\end{figure*}

\section{Related Work}
\label{sec:related_work}
In this section, we summarize past super-resolution (SR) techniques, explore diffusion-based models in SR, and address artifact detection and mitigation in the SR process.

\vspace{-4.0mm}
\paragraph{Image Super-Resolution.}
Single image super-resolution (SISR) is a process designed to construct high-resolution (HR) images from their low-resolution (LR) counterparts, which is inherently a complex task due to the multiple HR images that can correspond to a single LR image. 
The literature has seen a variety of solutions aiming to learn mappings from LR to HR images. These solutions fall into two primary groups: PSNR-oriented methods and generative model-based methods.
PSNR-oriented methods~\cite{dong2016accelerating,shi2016real,dong2015image,zhang2018residual,liang2021swinir,lim2017enhanced} employ L1 or L2 norms as objectives, achieving notable PSNR metrics. However, the reliance on L1/L2 losses tends to predict an average of all possible HR images~\cite{lugmayr2020srflow,ledig2017photo,johnson2016perceptual}, resulting in over-smoothed outputs.
In contrast, generative model-based methods~\cite{lugmayr2020srflow,wang2018esrgan,zhang2019ranksrgan,li2022srdiff,liang2021hierarchical,guo2022lar,ledig2017photo,lugmayr2022normalizing,bulat2018learn,zhang2022edface,wang2022survey} seek to address the ill-posed nature of the SR problem by learning the distribution of potential HR images through techniques such as GAN-based, diffusion-based, flow-based, and AR-based super-resolution. These models aim to generate HR images that not only have high pixel accuracy but also possess the rich textural details necessary for visual realism.

\vspace{-4.0mm}
\paragraph{Diffusion Models.}
Diffusion probabilistic models~\cite{sohl2015deep,ho2020denoising} are a class of generative models that utilize a Markov chain to convert latent variables from simple distributions (\eg, Gaussian) into data with complex distributions.
Recognizing the potential of these models to address ``one-to-many" mapping problems, Li~\etal~\cite{li2022srdiff} have applied them to the task of super-resolution, generating a variety of high-resolution images from a single low-resolution input. This application not only yields diverse outcomes but also simultaneously addresses challenges related to over smoothing, mode collapse, and the extensive computational footprint that are often encountered in super-resolution tasks.  
A prevalent method~\cite{saharia2022image,rombach2022high} in diffusion-based SR involves integrating a low-resolution image directly into the input of existing diffusion models (\eg, DDPM~\cite{ho2020denoising}), and retraining the model with super-resolution data from scratch.
Another approach~\cite{chung2022come,choi2021ilvr,yue2022difface} leverages an unconditional pre-trained diffusion model as a prior and adapts its reverse diffusion process to synthesize the desired high-resolution image. 
Nonetheless, both strategies are constrained by the intrinsic Markov chain sequence of DDPM, which can potentially lead to artifacts during inference, thus affecting the quality of the resulting high-resolution image.

\vspace{-4.0mm}
\paragraph{Artifact Detection Methods.}
Despite impressive improvement, diffusion-based super-resolution methods tend to present undesired outcomes, known as artifacts.
The artifact detection approach aims to probe the perceptually implausible contents rooted in the generated images.
Existing deep-learning-based image artifact detection methods can be divided into spatial and frequency domain-based approaches. 
The former approaches~\cite{yu2019attributing} focus on capturing differences in texture between real and generated images. Liu~\etal~\cite{liu2020global} leverage Gram matrices to capture the difference in texture between real and synthetic images.
Dang~\etal~\cite{dang2020detection} adopt the attention mechanism to improve the accuracy of artifact detection by highlighting informative regions.
Zhao~\etal~\cite{zhao2021multi} utilize the attention mechanism to amplify subtle differences in the shallow layers, improving the detection performance.
The latter approaches~\cite{zhang2019detecting} study the artifacts in the frequency domain.
Durall~\etal~\cite{durall2020watch} classify generated images according to the characteristics of the artifacts in high-frequency components.
Dzanic~\etal~\cite{dzanic2020fourier} demonstrate that image artifacts could be detected by leveraging the degree of partial decay at high frequencies.
Frank~\etal~\cite{dzanic2020fourier} find that the use of upsampling operations causes the artifacts in generated images, and employs the DCT transforms to detect them.

\section{Methodology}
\label{sec:method}
In this section, we propose a novel training-free method called "Self-Adaptive Reality-Guided Diffusion" (SARGD) to tackle the ongoing challenge of artifact suppression in diffusion-based super-resolution. This approach, depicted in Figure~\ref{fig:framework}, introduces two essential mechanisms.
\begin{itemize}
    \item  \textbf{Reality-Guided Refinement (RGR)}, mitigates artifact emergence during inference by employing guidance from a realistic latent space, thereby preserving the original characteristics of the images.
    \item  \textbf{Self-Adaptive Guidance (SAG)}, enhances the authenticity of the reference realistic latent representation, effectively addressing the over-smoothing tendencies associated with the initial realistic latent representation.
\end{itemize}

\subsection{Preliminaries}

Diffusion-based super-resolution (SR) techniques~\cite{li2022srdiff,saharia2022image,wang2023exploiting} have emerged as a compelling paradigm for the enhancement of image resolutions.
Consider the diffusion process as a function that aims to translate a given low-resolution (LR) image into a high-resolution (HR) image, each successive HR image at time $t+1$ is obtained by refining the current estimate, factoring in the original LR image. This refinement is mathematically represented as:
\begin{equation}
\small
\mathbf{I}_{HR}^{(t+1)} = \mathbf{I}_{HR}^{(t)} + f_\theta(\mathbf{I}_{HR}^{(t)}, \mathbf{I}_{LR}),
\end{equation}
where $\mathbf{I}_{HR}^{(t)}$ is the current state of the high-resolution image at iteration $t$.
$f_\theta$ is the diffusion function parameterized by $\theta$, which updates the current high-resolution state based on both its current state and the given low-resolution image. $t$ is the current iteration during the inference.

In the iterative diffusion process for super-resolution, the input LR-HR image pairs in the training set are used to train the diffusion models with the total diffusion step T.
The diffusion process initially leverages a pre-trained visual encoder to convert the upsampled LR image into a latent $z_t = \encoder(up(\mathbf{I}_{LR}))$, and gradually adding the Gaussian noise $\boldsymbol{\epsilon} \sim \mathcal{N}(\mathbf{0}, \boldsymbol{I})$ into it, resulting the noised latent variable $\boldsymbol{x}$.
The reverse diffusion process aims to guide the generation to the corresponding HR image through iterative refinement of the noised latent $\boldsymbol{x}$, taking the form of:
\begin{equation}
\small
    \boldsymbol{y}_{t-1} \leftarrow \frac{1}{\sqrt{\alpha_t}}\left(\boldsymbol{y}_t-\frac{1-\alpha_t}{\sqrt{1-\gamma_t}} f_\theta\left(\boldsymbol{x}, \boldsymbol{y}_t, \gamma_t\right)\right)+\sqrt{1-\alpha_t} \boldsymbol{\epsilon}_t, 
    \label{eq:reverse}
\end{equation}
where $\alpha_t \in (0,1)$ is the corresponding coefficient and $\gamma_t$ is the variance of the noise at iteration $t$. Lastly, the decoder $\decoder$ is employed to project the resulting latent into the final high-resolution image.

\begin{figure}
    \centering
    \includegraphics[width=\linewidth]{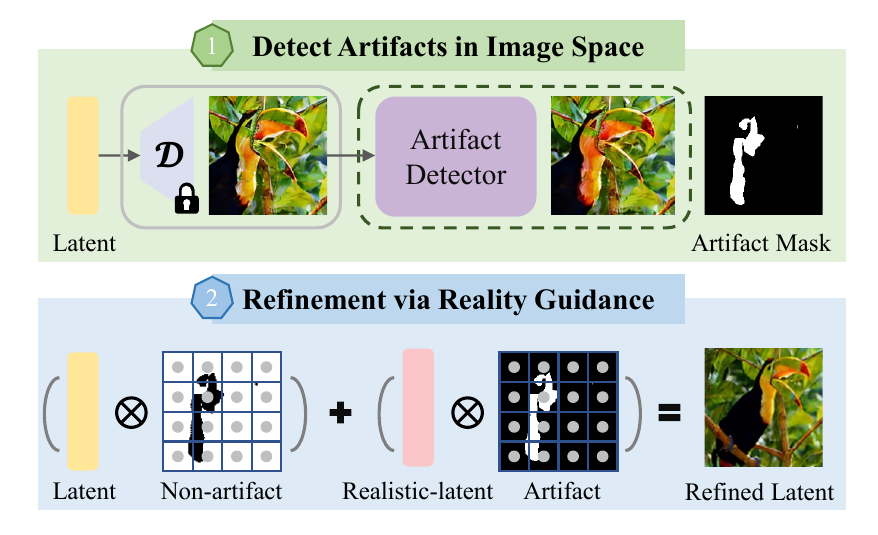}
    \vspace{-8.0mm}
    \caption{\textbf{Overview of the Reality-Guided Refinement (RGR) workflow}: 1) Decoding the current latent into an RGB image and using an artifact detector to create a binary mask identifying areas with artifacts; and 2) Utilizing realistic latent guidance to refine the masked regions, enhancing the image's fidelity and authenticity.}
    \vspace{-3.0mm}
    \label{fig:RGD}
\end{figure}

\subsection{Reality-Guided Refinement (RGR)}

To overcome the prevalent dilemma of aggravated noise and the emergence of artifacts in images subjected to super-resolution (SR) via diffusion-based techniques, we propose a pioneering training-free approach: Reality-Guided Refinement (RGR).
This method couples an artifact detector with a latent refinement to address these issues effectively.
As illustrated in Figure~\ref{fig:RGD}, a pixel-level segmentation model is initially applied to locate any artifacts on the current latent variable.
Once detected, these imperfections are adeptly rectified through a sophisticated procedure of realistic-latent guidance refinement.
This diffusion process ensures that the resulting images retain the original, authentic textures and intricate details, effectively circumventing the usual distortions and artifacts associated with traditional diffusion-based upsampling techniques for SR tasks.

\vspace{-4.0mm}
\paragraph{Artifact Detection in Diffusion Process.} 
In the context of pixel-sensitive super-resolution, it is crucial to identify the fine-grained artifacts during the diffusion process.
Hence, a robust artifact detection system should precisely pinpoint these implausible pixels within the latent space for effective identification.
To achieve this, we utilize the established artifact detection method named PAL~\cite{Zhang_2023_ICCV} to isolate these imperfections in the latent space.
The process entails transforming the latent variable into the image domain at each iteration using the decoder, depicted by the following equation:
\begin{equation}
\small
\mathbf{I}_{HR}^{(t-1)} = \decoder(\frac{1}{\sqrt{\alpha_t}}\left(\boldsymbol{x}_t-\frac{1-\alpha_t}{\sqrt{1-\bar\alpha_t}} \epsilon_\theta(\boldsymbol{x_t},\boldsymbol{x},t)\right) + \sigma_\theta(x_t, t) \boldsymbol{\epsilon})),
\end{equation}
Subsequently, the resulting image is input into the PAL detector,  yielding a binary mask  $E_A$ that denotes the presence of artifacts (referenced as line 12 in Algorithm~\ref{alg:realtiy_guided_diffusion}). 
This process can be formulated as:
\begin{equation}
\small
E_A^{(t-1)} = \mathcal{A}(\mathbf{I}_{HR}^{(t-1)}).
\end{equation}
Finally, the resultant artifact mask is resized to match the scale of the latent variable, maintaining the proportionality between the two.

\vspace{-4.0mm}
\paragraph{Refinement via Reality Guidance.}
For the removal of artifacts during sampling, we refine the latent variable via a proposed training-free RGR. 
Following Algorithm~\ref{alg:realtiy_guided_diffusion}, the RGR inference, consisting of T steps, begins by taking an upscaled low-resolution image $up(\mathbf{I}_{LR})$ as the starting point. 
A noise vector $\boldsymbol{\epsilon}$ is then drawn from a standard Gaussian distribution. 
The process iterates backward from $t=T$, with each step generating a latent variable $\boldsymbol{x}_{t-1}$ in accordance with
DDPM~\cite{ho2020denoising} framework.
Once artifacts $E_A$ are identified within the current latent $\boldsymbol{x}_{t-1}$, we proceed to refine these imperfections through reality guidance. This correction process unfolds as follows:
\begin{equation}
\small
    \boldsymbol{x}_{t-1} = \boldsymbol{x}_{t-1} \times (1-E_A) + \boldsymbol{x}_r \times E_A.
\end{equation}
This process allows us to effectively eliminate artifacts from the latent space.
Notably, in this formula, we employ the initially encoded latent, which has not undergone the denoising process, as a source of realistic latent guidance, since it retains the intrinsic structural integrity of the image.

\subsection{Self-Adaptive Guidance (SAG)}
However, relying on the initial latent space as a guide for artifact suppression during inference often results in over-smoothing results, primarily attributed to reliance on upscaled low-resolution images.
To address this, we propose a novel Self-Adaptive Guidance (SAG) renewal mechanism, an automated system that perpetually refines realistic latent representations, ensuring the guidance is continually calibrated to enhance realistic detail.
This mechanism is indispensable for diffusion-based super-resolution, aiming to yield high-resolution and high-fidelity images that strike a balance between details and sharpness. 

\vspace{-4.0mm}
\paragraph{Reality Score Computation.} 
To regulate the update mechanism of reality guidance in image super-resolution, we introduce an innovative metric known as the reality score. This metric quantifies the realism of latent guidance. 
As delineated in Algorithm~\ref{alg:realtiy_guided_diffusion}, we begin with an artifact-free latent variable $\boldsymbol{x}_{t-1}$and utilize our decoding function to synthesize a realistic image $\mathbf{I}_{r} = \decoder(\boldsymbol{x}_{t-1})$. 
Subsequently, we deploy a realistic-region function $\mathcal{R}$ by inverting the output of PAL~\cite{Zhang_2023_ICCV} on $\mathbf{I}_{r}$, resulting in a binary reality mask $M_R$ with dimensions corresponding to the channels, height, and width of the latent $C \times H \times W$, respectively.
The reality score $\boldsymbol{s}_r$ at the current $t-1$ step is calculated as the normalized sum of the reality mask across the spatial dimensions of the latent, indicating the distribution of realistic features. The formula is given by:
\begin{equation}
\small
    s_r^{t-1} = \mathcal{S}(M_R) = \frac{1}{H \times W}\sum_{i=0}^{H-1}\sum_{j=0}^{W-1} M_R^{(ij)}.
\end{equation}
Herein, \(H\) and \(W\) are the height and width of the latent, and \(M_R^{(ij)}\) indicates the reality measure at each pixel. 
This approach ensures a quantitative evaluation of the SR process, focusing on producing images that boast both high resolution and a high degree of fidelity to reality.

\vspace{-4.0mm}
\paragraph{Self-Adaptive Reality Guidance.}
Self-adaptability in the realistic latent space refers to the ability of the system to adjust the reality guidance to correct the details and textures that are characteristic of high-quality images.
This is especially important to mitigate the smoothing effect of the initial realistic-latent reliance on bicubic upsampling.
As outlined in Algorithm~\ref{alg:realtiy_guided_diffusion}, we begin by considering a realistic image $\mathbf{I}_r$ at the current step, decoded from an artifact-free latent variable $\boldsymbol{x}_{t-1}$.
This artifact-free image is then encoded into a realistic latent  $\boldsymbol{x}_{r}^{t-1}=\encoder(\mathbf{I}_r)$.
Grounded on the computed reality score, self-adaptability is characterized as a guidance refinement process. 
During this process, the realistic latent $\boldsymbol{x}_r$ at the current $t-1$ step is iteratively refined to converge towards an ideal latent representation $\boldsymbol{x}r$, precisely aligning with the high-resolution output $\mathbf{I}{HR}$. The equation below illustrates this adaptive mechanism:
\begin{equation}
\small
\mathcal{G}(\boldsymbol{x}_{r}, \boldsymbol{x}_r^{t-1}) = \boldsymbol{x}_r \times (1 - M_R) + \boldsymbol{x}_{r}^{t-1}  \times M_R,  \ \  \text{if } \boldsymbol{s}_r^{t-1} > \boldsymbol{s}_r, 
\end{equation}
where $\boldsymbol{s}_r$ represents the reality score associated with $\boldsymbol{x}r$, and $M_R$ is the binary reality mask corresponding to $\boldsymbol{x}{r}^{t-1}$. This self-adaptive mechanism ensures that the guidance is autonomously updated, leading to enhanced super-resolution outcomes.

\setcounter{algorithm}{0}
\begin{algorithm}[t] 
    \centering 
    \small
    \caption{Self-Adaptive Reality-Guided Diffusion}
    \label{alg:realtiy_guided_diffusion} 
    \begin{algorithmic}[1] 
        \STATE\textbf{Input}: LR image $\mathbf{I}_{LR}$, and total diffusion steps $T$
        \STATE\textbf{Load}: Encoder $\encoder$, artifact detector $\mathcal{A}$ and LR decoder $\decoder$
        \STATE $\blacktriangleright$ \textbf{Step 1: Initialization}
        \STATE Upscale LR image as $up(\mathbf{I}_{LR})$
        \STATE Encode the upsampled image as $\boldsymbol{x} = \encoder
        (up(\mathbf{I}_{LR}))$ 
        \STATE Initialize the $\boldsymbol{x}$ as a realistic latent $\boldsymbol{x}_{r}$ and set it as guidance
        \STATE Compute the realty score of the realistic latent $\boldsymbol{s}_{r}$
        \STATE $\blacktriangleright$ \textbf{Step 2: Sampling }
        \FOR{$t=T,\cdots,1$} 
            \STATE Sample $\boldsymbol{\epsilon} \sim \mathcal{N}(\textbf{0}, \textbf{I})$ if $t > 1$, else $\boldsymbol{\epsilon} = 0$ 
            \STATE Compute the latent variable at the current step $\boldsymbol{x}_{t-1} =\frac{1}{\sqrt{\alpha_t}}\left(\boldsymbol{x}_t-\frac{1-\alpha_t}{\sqrt{1-\bar\alpha_t}} \epsilon_\theta(\boldsymbol{x_t},\boldsymbol{x},t)\right) + \sigma_\theta(\boldsymbol{x}_t, t) \boldsymbol{\epsilon}$
            \STATE \textcolor{myred}{$\bigstar$ \textbf{Reality-Guided Refinement}}
            \STATE \textbf{Detect artifacts} of the current latent $E_A = \mathcal{A}(\decoder(\boldsymbol{x}_{t-1}))$
            \STATE \textbf{Refine the latent} $\boldsymbol{x}_{t-1} = \boldsymbol{x}_{t-1} \times (1-E_A) + \boldsymbol{x}_r \times E_A$
            \STATE \textcolor{myblue}{$\bigstar$ \textbf{Reality Score Computation}}
            \STATE Decode the refined latent into an image $\mathbf{I}_r = \decoder(\boldsymbol{x}_{t-1})$
            \STATE Generate the current binary reality map $M_R = \mathcal{R}(\mathbf{I}_r)$ 
            \STATE Calculate the current reality score $\boldsymbol{s}_{r}^{t-1} = \mathcal{S}(M_R)$
            \STATE \textcolor{mygreen}{$\bigstar$ \textbf{Self-Adaptive Guidance}}
            \STATE \textbf{Encode the current realistic latent} $\boldsymbol{x}_r^{t-1}=\encoder(\mathbf{I}_r)$
            \STATE \textbf{Update the guidance} $\boldsymbol{x}_{r} = \mathcal{G}(\boldsymbol{x}_{r}, \boldsymbol{x}_r^{t-1})$ if $\boldsymbol{s}_{r}^{t-1} > \boldsymbol{s}_{r}$
            \STATE \textbf{Update the reality score} $\boldsymbol{s}_{r} = \boldsymbol{s}_{r}^{t-1}$ if $\boldsymbol{s}_{r}^{t-1} > \boldsymbol{s}_{r}$
        \ENDFOR 
        \RETURN the artifact-free SR $\mathbf{L}_{HR} = \decoder(\boldsymbol{x}_0)$
    \end{algorithmic} 
\end{algorithm}

\begin{table*}
    \centering
    \caption{\textbf{Quantitative comparison with the state-of-the-art \underline{Diffusion-SR} methods across benchmark datasets for super-resolution at $\times 2$, $\times 3$, and $\times 4$ scales.} The dagger symbol ($\dag$) signifies LDM trained on the identical dataset as used for StableSR. \textbf{Bold} highlights the best performance. All evaluations are conducted on a 32G GPU. Our training-free SARGD attains the most favorable results. \vspace{-3.0mm}}
    \label{tab:comp_benchmark}
    \begin{adjustbox}{max width=\textwidth}
    \begin{tabular}{r|ccccccccccccccccccccc}
    \Xhline{1.5pt}
    \rowcolor{mygray}
     \multicolumn{2}{c}{\textbf{Benchmarks}}&
    &\multicolumn{3}{c}{\textbf{Set5}~\cite{bevilacqua2012low}}&
    &\multicolumn{3}{c}{\textbf{Set14}~\cite{zeyde2012single}}&
    &\multicolumn{3}{c}{\textbf{B100}~\cite{martin2001database}}&
    &\multicolumn{3}{c}{\textbf{Urban100}~\cite{huang2015single}}&
    &\multicolumn{3}{c}{\textbf{Manga109}~\cite{matsui2017sketch}} \\
    \Xcline{1-2}{1.0pt} \Xcline{4-6}{1.0pt} \Xcline{8-10}{1.0pt} \Xcline{12-14}{1.0pt} \Xcline{16-18}{1.0pt} \Xcline{20-22}{1.0pt}
    \rowcolor{mygray}
    Metrics &Methods & & $\times2$ & $\times3$ & $\times4$ & & $\times2$ & $\times3$ & $\times4$ & & $\times2$ & $\times3$ & $\times4$ & & $\times2$ & $\times3$ & $\times4$ & & $\times2$ & $\times3$ & $\times4$ \\
    \hline \toprule
    
    \multirow{4}{*}{PSNR $\uparrow$ } 
                                 &LDM$^\dag$    &   &27.86 &27.94 &28.30
                                                &   &26.42 &26.81 &27.21
                                                &   &26.91 &27.54 &27.67
                                                &   &25.18 &25.72 &25.91
                                                &   &26.49 &26.96 &27.04 \\
                                     &StableSR  &   &28.78 &29.09 &29.36
                                                &   &27.26 &27.69 &28.04
                                                &   &27.61 &28.28 &28.40
                                                &   &26.00 &26.44 &26.55
                                                &   &27.58 &28.01 &27.94 \\
                                      &Ours     &   &\textbf{32.64} &\textbf{33.06} &\textbf{32.27}
                                                &   &\textbf{31.40} &\textbf{30.94} &\textbf{30.01}
                                                &   &\textbf{30.92} &\textbf{30.84} &\textbf{30.23}
                                                &   &\textbf{29.34} &\textbf{28.84} &\textbf{27.93}
                                                &   &\textbf{31.21} &\textbf{31.13} &\textbf{30.23} \\
    \cmidrule{1-22}
    \multirow{4}{*}{SSIM $\uparrow$ }    
                                &LDM$^\dag$     &   &0.765 &0.777 &0.769
                                                &   &0.673 &0.688 &0.696
                                                &   &0.651 &0.672 &0.679
                                                &   &0.712 &0.719 &0.719
                                                &   &0.814 &0.821 &0.815 \\
                                     &StableSR  &   &0.805 &0.814 &0.812
                                                &   &0.716 &0.731 &0.733
                                                &   &0.694 &0.714 &0.717
                                                &   &0.745 &0.748 &0.744
                                                &   &0.843 &0.847 &0.838 \\
                                      &Ours     &   &\textbf{0.899} &\textbf{0.897} &\textbf{0.871}
                                                &   &\textbf{0.842} &\textbf{0.823} &\textbf{0.778}
                                                &   &\textbf{0.817} &\textbf{0.800} &\textbf{0.763}
                                                &   &\textbf{0.849} &\textbf{0.817} &\textbf{0.771}
                                                &   &\textbf{0.909} &\textbf{0.897} &\textbf{0.851} \\
    \cmidrule{1-22}
    \multirow{4}{*}{LPIPS $\downarrow$ }   
                                &LDM$^\dag$     &   &0.176 &0.172 &0.163
                                                &   &0.217 &0.214 &0.236
                                                &   &0.222 &0.238 &0.252
                                                &   &0.167 &0.176 &0.186
                                                &   &0.132 &0.128 &0.138 \\
                                     &StableSR  &   &0.165 &0.153 &0.150
                                                &   &0.197 &0.194 &0.207
                                                &   &0.200 &0.214 &0.226
                                                &   &0.150 &0.159 &\textbf{0.166}
                                                &   &0.118 &0.115 &\textbf{0.123} \\
                                      &Ours     &   &\textbf{0.118} &\textbf{0.126} &\textbf{0.131}
                                                &   &\textbf{0.139} &\textbf{0.149} &\textbf{0.204}
                                                &   &\textbf{0.164} &\textbf{0.118} &\textbf{0.224}
                                                &   &\textbf{0.117} &\textbf{0.138} &0.177
                                                &   &\textbf{0.083} &\textbf{0.085} &0.128 \\
    \cmidrule{1-22}
    \multirow{4}{*}{DISTS $\downarrow$ } 
                                &LDM$^\dag$     &   &0.172 &0.173 &0.161
                                                &   &0.157 &0.158 &0.162
                                                &   &0.167 &0.171 &0.176
                                                &   &0.129 &0.134 &0.141
                                                &   &0.102 &0.098 &0.103 \\
                                     &StableSR  &   &0.167 &0.165 &0.157
                                                &   &0.152 &0.151 &\textbf{0.150}
                                                &   &0.160 &0.160 &0.164
                                                &   &0.127 &0.128 &\textbf{0.131}
                                                &   &0.102 &0.097 &0.099 \\
                                      &Ours     &   &\textbf{0.136} &\textbf{0.134} &\textbf{0.142}
                                                &   &\textbf{0.126} &\textbf{0.121} &0.156
                                                &   &\textbf{0.132} &\textbf{0.140} &\textbf{0.156}
                                                &   &\textbf{0.111} &\textbf{0.119} &0.139
                                                &   &\textbf{0.086} &\textbf{0.083} &\textbf{0.096} \\

    \bottomrule\hline 
    \end{tabular}
    \end{adjustbox}
\end{table*}

\section{Experimental Results}
\label{sec:experiments}

In this section, we present our experimental results, discuss insights from ablation studies, and delve into the broader implications of our findings.

\subsection{Experimental Setups} 

\paragraph{Testing Datasets.} Our proposed SARGD method is rigorously assessed for arbitrary-scale single image super-resolution (SR) across a comprehensive set of benchmark datasets, including Set5~\cite{bevilacqua2012low}, Set14~\cite{zeyde2012single}, B100~\cite{martin2001database}, Urban100~\cite{huang2015single}, and Manga109~\cite{matsui2017sketch}. 
The Set5 and Set14 datasets serve as standard benchmarks within the super-resolution domain, comprising five and fourteen high-resolution images, respectively. 
The B100 dataset is a collection of 100 natural images used for testing the versatility of super-resolution algorithms across various real-world scenarios.
Urban100 comprises 100 detailed urban images, challenging super-resolution algorithms to accurately enhance intricate structures.
Manga109 contains 109 high-resolution manga images, testing the edge preservation capabilities of super-resolution models.

\begin{figure*}
    \centering
    \includegraphics[width=\textwidth]{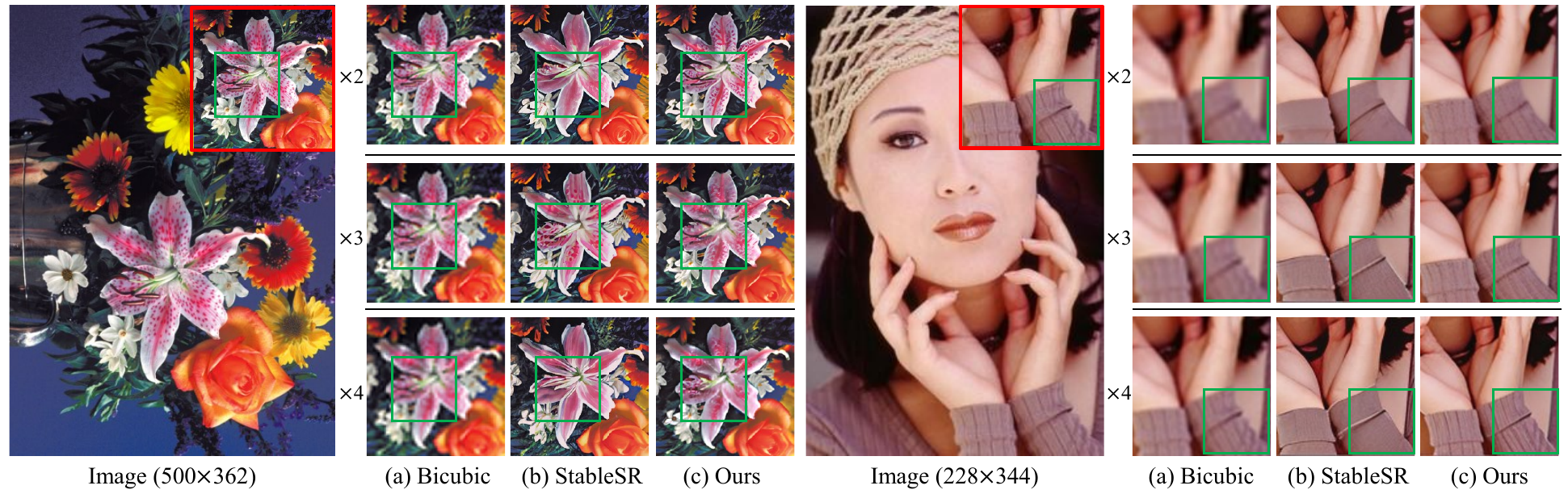}
    \vspace{-7.0mm}
    \caption{\textbf{Visual comparison with diffusion-SR methods} for $\times 2$, $\times 3$, and $\times 4$ super-resolution, including (a) \textbf{Bicubic upsampling}, (b) \textbf{StableSR}, and (c) our \textbf{\textit{SARGD}}. The {\color{myred}{red}} solid-lined boxes represent the ground truth (GT), focusing on regions zoomed for detailed inspection. The {\color{mygreen}{green}} boxes illustrate how our SARGD method preserves significantly more detail and clarity compared to the alternatives.}
    \vspace{-5.0mm}
    \label{fig:benchmark_viz}
\end{figure*}

\vspace{-4.0mm}
\paragraph{Evaluation Metrics.} 
We utilize established perceptual metrics including PSNR, SSIM~\cite{wang2004image}, LPIPS~\cite{zhang2018unreasonable}, and DISTS~\cite{zhang2018image}
to conduct a comprehensive comparison with existing diffusion-based SR approaches.
PSNR and SSIM, particularly when assessed on the Y channel of the YCbCr color space, are fidelity-focused measures traditionally favored in super-resolution tasks. Typically, higher values of PSNR and SSIM correspond to better performance. 
Nevertheless, these metrics do not adequately assess the quality of generative models.
Therefore, we also incorporate LPIPS and DISTS in our assessment suite, as they have been acknowledged for their effectiveness in evaluating the perceptual quality of images produced by generative models.
Generally, lower scores in LPIPS and DISTS are indicative of superior perceptual image quality.

\vspace{-4.0mm}
\paragraph{Implementation Details.} 
We adhere to the experimental protocol established by previous research~\cite{wang2023exploiting}. 
All of our experiments are conducted using PyTorch and tested on a consistent computing platform.
The uniform model trained by Zhang~\etal~\cite{Zhang_2023_ICCV} is employed as an artifact detector to generate the artifact mask and compute the reality score. 
For the inference phase, we set the number of denoising steps to 200.
Following established protocols~\cite{chen2021learning}, we derive low-resolution test samples using the imresize function in Matlab, employing bicubic interpolation with downsampling scales of $\times2$, $\times3$, and $\times4$, in line with the precedents set by prior research. 
In diffusion-based super-resolution methods, it is standard practice to upsample the inputs to the same size as the outputs prior to the inference stage.

\begin{table*}
    \centering
    \caption{\textbf{Analysis of SARGD inference strategies.}
    The `Baseline' model represents the absence of Reality-Guided Refinement (RGR) and Self-Adaptive Guidance (SAG). `RGR' utilizes initial realistic latent guidance, while `SAG' employs the self-adaptive guidance mechanism. 
    All experiments are conducted under identical settings. The best scores are highlighted in \textbf{bold}.
    }
    \vspace{-3.0mm}
    \label{tab:sargd_ablation_study}
    \begin{adjustbox}{max width=\textwidth}
    \begin{tabular}{r|cc|ccccccccccccccc}
    \Xhline{1.5pt}
    \rowcolor{mygray}
    & &
    &\multicolumn{3}{c}{\textbf{Set5}~\cite{bevilacqua2012low}}&
    &\multicolumn{3}{c}{\textbf{Set14}~\cite{zeyde2012single}}&
    &\multicolumn{3}{c}{\textbf{B100}~\cite{martin2001database}}&
    &\multicolumn{3}{c}{\textbf{Manga109}~\cite{matsui2017sketch}} \\
    \Xcline{4-6}{1.0pt} \Xcline{8-10}{1.0pt} \Xcline{12-14}{1.0pt} \Xcline{16-18}{1.0pt}
    \rowcolor{mygray}
      \multirow{-2}{*}{\textbf{Metric}} 
    & \multirow{-2}{*}{\textbf{RGR}} 
    &\multirow{-2}{*}{\textbf{SAG}} 
    &$\times2$ &$\times3$ &$\times4$& 
    &$\times2$ &$\times3$ &$\times4$& 
    &$\times2$ &$\times3$ &$\times4$& 
    &$\times2$ &$\times3$ &$\times4$ \\
    \hline \hline
    
     \multirow{3}{*}{PSNR $\uparrow$} 
     &\XSolidBrush &\XSolidBrush &27.86 &27.94 &28.30 &  &26.42 &26.81 &27.21 &  &26.91 &27.54 &27.67 &  &26.49 &26.96 &27.04 \\
     &\color{myred}\Checkmark   &\XSolidBrush 
        &29.82 &30.39 &30.27 {(\textcolor{mygreen}{+ 6.96})}
     &  &28.16 &27.99 &27.79 {(\textcolor{mygreen}{+ 2.13})}
     &  &28.08 &28.42 &28.30 {(\textcolor{mygreen}{+ 2.28})}
     &  &25.10 &26.41 &27.88 {(\textcolor{mygreen}{+ 3.11})} \\    
     &\color{myred}\Checkmark   &\color{mygreen}\Checkmark   
        &\textbf{32.42} &\textbf{32.93} &\textbf{31.79} {(\textcolor{mygreen}{+ 12.3})}
     &  &\textbf{30.16} &\textbf{30.33} &\textbf{29.28} {(\textcolor{mygreen}{+ 7.61})}
     &  &\textbf{30.29} &\textbf{30.38} &\textbf{29.92} {(\textcolor{mygreen}{+ 8.13})}
     &  &\textbf{30.40} &\textbf{30.58} &\textbf{29.23} {(\textcolor{mygreen}{+ 8.10})}\\   
     \cmidrule{1-18}
 
     \multirow{3}{*}{SSIM $\uparrow$}
     &\XSolidBrush &\XSolidBrush &0.765 &0.777 &0.769 &  &0.673 &0.688 &0.696 &  &0.651 &0.672 &0.679 &  &0.814 &0.821 &0.815 \\
     &\color{myred}\Checkmark   &\XSolidBrush 
       &0.823 &0.843 &0.824 {(\textcolor{mygreen}{+ 7.15})}
     & &0.728 &0.732 &0.730 {(\textcolor{mygreen}{+ 4.89})}
     & &0.702 &0.711 &0.704 {(\textcolor{mygreen}{+ 3.68})}
     & &0.775 &0.808 &0.822 {(\textcolor{mygreen}{+ 0.86})} \\
     &\color{myred}\Checkmark   &\color{mygreen}\Checkmark   
        &\textbf{0.897} &\textbf{0.895} &\textbf{0.870} {(\textcolor{mygreen}{+ 13.1})} 
     &  &\textbf{0.821} &\textbf{0.815} &\textbf{0.770} {(\textcolor{mygreen}{+ 10.6})} 
     &  &\textbf{0.807} &\textbf{0.792} &\textbf{0.760} {(\textcolor{mygreen}{+ 11.9})}
     &  &\textbf{0.900} &\textbf{0.893} &\textbf{0.851} {(\textcolor{mygreen}{+ 4.41})}\\ 
     \cmidrule{1-18}
    
     \multirow{3}{*}{LPIPS $\downarrow$}
     &\XSolidBrush &\XSolidBrush &0.176 &0.172 &0.163 &  &0.217 &0.214 &0.236 &  &0.222 &0.238 &0.252 &  &0.132 &0.128 &0.138 \\
     &\color{myred}\Checkmark   &\XSolidBrush 
        &0.144 &0.212 &0.161 {(\textcolor{mygreen}{- 1.23})}
     &  &0.227 &0.270 &0.242 {(\textcolor{color1}{+ 2.54})}
     &  &0.201 &0.248 &0.292 {(\textcolor{color1}{+ 15.8})}
     &  &0.128 &0.139 &0.164 {(\textcolor{color1}{+ 18.9})}\\ 
     &\color{myred}\Checkmark   &\color{mygreen}\Checkmark   
        &\textbf{0.122} &\textbf{0.141} &\textbf{0.131} {(\textcolor{mygreen}{- 19.6})} 
     &  &\textbf{0.144} &\textbf{0.152} &\textbf{0.208} {(\textcolor{mygreen}{- 11.9})}
     &  &\textbf{0.172} &\textbf{0.201} &\textbf{0.228} {(\textcolor{mygreen}{- 9.52})}
     &  &\textbf{0.087} &\textbf{0.087} &\textbf{0.132} {(\textcolor{mygreen}{- 4.35})}\\
     \cmidrule{1-18}
    
     \multirow{3}{*}{DISTS $\downarrow$} 
     &\XSolidBrush &\XSolidBrush &0.172 &0.173 &0.161 &  &0.157 &0.158 &0.162 &  &0.167 &0.171 &0.176 &  &0.102 &0.098 &0.103 \\
     &\color{myred}\Checkmark   &\XSolidBrush 
        &0.145 &0.173 &0.192 {(\textcolor{color1}{+ 19.3})} 
     &  &0.160 &0.179 &0.214 {(\textcolor{color1}{+ 32.1})} 
     &  &0.191 &0.213 &0.238 {(\textcolor{color1}{+ 35.2})} 
     &  &0.094 &0.098 &0.111 {(\textcolor{color1}{+ 7.77})} \\
     &\color{myred}\Checkmark   &\color{mygreen}\Checkmark   
        &\textbf{0.137} &\textbf{0.141} &\textbf{0.145} {(\textcolor{mygreen}{- 9.94})} 
     &  &\textbf{0.127} &\textbf{0.129} &\textbf{0.158} {(\textcolor{mygreen}{- 2.47})} 
     &  &\textbf{0.138} &\textbf{0.141} &\textbf{0.158} {(\textcolor{mygreen}{- 10.2})} 
     &  &\textbf{0.091} &\textbf{0.084} &\textbf{0.099} {(\textcolor{mygreen}{- 3.88})} \\                                       
    \Xhline{1.5pt}
    \end{tabular}
    \end{adjustbox}
\vspace{-1.0em}
\end{table*}

\subsection{Comparison with State-of-the-Art}
To verify the effectiveness of our training-free SARGD, we conducted comparative analyses with various state-of-the-art methods, such as LDM~\cite{rombach2022high} and StableSR~\cite{wang2023exploiting}. 
To ensure a fair comparison, LDM is finetuned using the same training configurations and dataset as those employed for StableSR.
For other methods, we directly use the official codes and pre-trained models for testing. 
Comprehensive results, along with additional comparative analyses, are available in the supplementary material.

\vspace{-4.0mm}
\paragraph{Quantitative Results.}
Table~\ref{tab:comp_benchmark} demonstrates the outstanding capability of our proposed Self-Adaptive Reality-Guided Diffusion (SARGD) method in producing artifact-free super-resolution images.
Notably, our method significantly surpasses the previous StableSR model in terms of PSNR, with improvements of 3.86, 3.97, and 2.91 on the Set5 dataset at upscaling factors of $\times$2, $\times$3, and $\times$4, respectively. 
Within the SSIM metric, our SARGD method realizes significant improvements (\eg the increase from 0.716 to 0.842 on Set14 and from 0.694 to 0.817 on B100), illustrating a substantial enhancement in structural fidelity.
Additionally, in terms of the DISTS metric, SARGD significantly reduces the score from 0.102 to 0.086 in Manga109, reflecting an improved alignment with human perceptual quality.
These improvements establish our SARGD as a leading approach in the field of super-resolution, setting a new standard for image clarity and detail preservation.

\begin{figure}
    \centering
    \includegraphics[width=\linewidth]{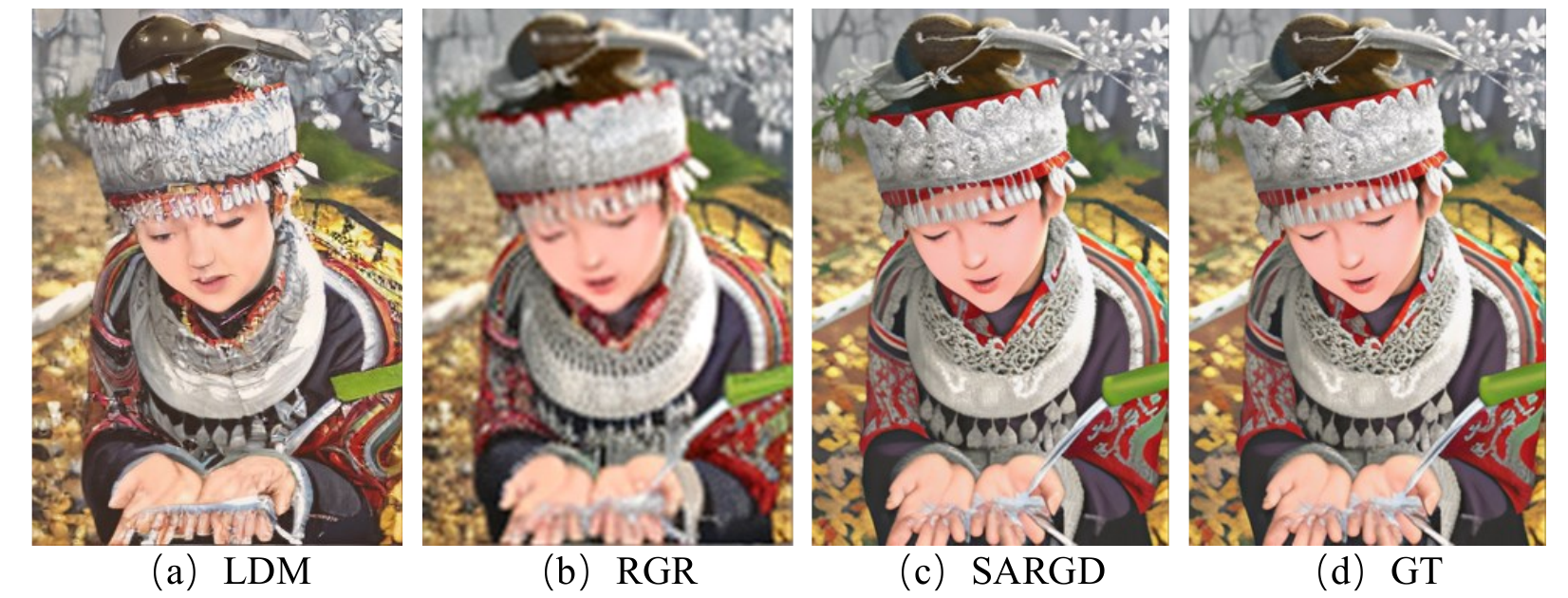}
    \vspace{-7.0mm}
    \caption{\textbf{Visual comparison of SARGD components.} Our SARGD exhibits the best outcomes for $\times 3$ super-resolution.}
    \vspace{-3.0mm}
    \label{fig:fixed_and_adaptive_latent}
\end{figure}

\vspace{-4.0mm}
\paragraph{Visualization.}
Figure~\ref{fig:benchmark_viz} offers a visual comparison between our SARGD and StableSR~\cite{wang2023exploiting}, further highlighting the superior performance of our method.
Compared to columns (a) and (c), bicubic upscaling maintains satisfactory detail but often leads to an over-smoothing issue.
While StableSR succeeds in generating high-resolution images, its dependence on conventional DDPM methods makes it susceptible to introducing additional artifacts.
When compared to columns (b) and (c), our approach successfully produces images with high authenticity, thereby avoiding the over-smoothing problem commonly seen in other methods.

\subsection{Ablation Study}
In this section, we thoroughly evaluate and analyze the performance of the proposed training-free SARGD method, primarily focusing on its effectiveness in denoising.

\begin{table}
    \centering
    \caption{\textbf{Comparison of varied denoising approaches within RGR} across Set14, B100, and Urban100 datasets for $\times$2, $\times$3, $\times$4 upscaling. `$\mathbf{D}_{1s}^{*}$' refers to artifact refinement in the latent space at each step, `$\mathbf{D}_{10s}^{*}$' at every ten steps, while `$\mathbf{D}_{100s}^{-}$' and `$\mathbf{D}_{100s}^{+}$' indicate refinement before and after 100 steps, respectively.}
    \vspace{-3.0mm}
    \label{tab:guided_strategy_rgr}
    \begin{adjustbox}{max width=\linewidth}
    \begin{tabular}{c|l|cc|cc|cc}
    \Xhline{1.5pt}
    \rowcolor{mygray}
    &
    &\multicolumn{2}{c|}{\textbf{Set14}}
    &\multicolumn{2}{c|}{\textbf{B100}}
    &\multicolumn{2}{c}{\textbf{Urban100}}\\
    \Xcline{3-4}{1.0pt} \Xcline{5-6}{1.0pt} \Xcline{7-8}{1.0pt}
    \rowcolor{mygray}
     \multirow{-2}{*}{\textbf{Scale}}
    &\multirow{-2}{*}{\textbf{Type}}
    &\textbf{PSNR $\uparrow$}
    &\textbf{SSIM $\uparrow$}
    &\textbf{PSNR $\uparrow$}
    &\textbf{SSIM $\uparrow$}
    &\textbf{PSNR $\uparrow$}
    &\textbf{SSIM $\uparrow$} \\
    \hline \hline 
    \multirow{4}{*}{$\times 2$} &$\mathbf{D}_{1s}^{*}$    &28.16 &0.728 &28.08 &0.702 &25.28 &0.708 \\
                                &$\mathbf{D}_{10s}^{*}$   &27.99 &0.725 &28.05 &0.692 &25.11 &0.703 \\
                                &$\mathbf{D}_{100s}^{-}$  &26.79 &0.685 &27.32 &0.672 &24.42 &0.676\\
                                &$\mathbf{D}_{100s}^{+}$  &\textbf{28.42} &\textbf{0.739} &\textbf{28.11} &\textbf{0.715} &\textbf{25.61} &\textbf{0.716} \\
    \midrule
    \multirow{4}{*}{$\times 3$} &$\mathbf{D}_{1s}^{*}$    &27.99 &0.732 &28.42 &0.711 &26.25 &0.751 \\
                                &$\mathbf{D}_{10s}^{*}$   &27.83 &0.726 &28.47 &0.705 &26.14 &0.747 \\
                                &$\mathbf{D}_{100s}^{-}$  &26.64 &0.686 &27.66 &0.688 &25.78 &0.731 \\
                                &$\mathbf{D}_{100s}^{+}$  &\textbf{28.02} &\textbf{0.730} &\textbf{28.54} &\textbf{0.718} &\textbf{26.59} &\textbf{0.759} \\
    \midrule
    \multirow{4}{*}{$\times 4$} &$\mathbf{D}_{1s}^{*}$    &27.79 &0.730 &28.30 &0.704 &25.92 &0.711\\
                                &$\mathbf{D}_{10s}^{*}$   &27.83 &0.729 &28.48 &0.702 &25.88 &0.708\\
                                &$\mathbf{D}_{100s}^{-}$  &26.94 &0.708 &27.83 &0.689 &25.47 &0.702\\
                                &$\mathbf{D}_{100s}^{+}$  &\textbf{28.33} &\textbf{0.734} &\textbf{28.51} &\textbf{0.714} &\textbf{25.96} &\textbf{0.713} \\
    \Xhline{1.5pt}
    \end{tabular}
    \end{adjustbox}
\vspace{-1.0em}
\end{table}

\vspace{-4.0mm}
\paragraph{Analysis of Improvement.}
Table~\ref{tab:sargd_ablation_study} presents the effects of our proposed modules, Reality-Guided Refinement (RGR) and Self-Adaptive Guidance (SAG).
The baseline corresponds to the LDM configured and trained similarly to StableSR~\cite{wang2023exploiting}.  
Analysis of the initial two columns reveals that the integration of RGR notably enhances performance, with gains of 6.96\%, 2.13\%, 2.28\%, and 3.11\% observed in $\times$4 super-resolution on the Set5, Set14, B100, and Manga109 datasets, respectively. 
The integration of RGR into the DDPM architecture leads to an increase in SSIM scores, validating the effectiveness of RGR in enhancing image quality.
The visual results in Figure~\ref{fig:fixed_and_adaptive_latent} show that while RGR manages to maintain the essential attributes of the original images, it also leads to over-smoothing in the super-resolved outputs.
By contrasting figures (b) and (c), it is evident that incorporating our SAG during inference enhances image quality and successfully addresses the issue of over-smoothing, thus confirming the effectiveness of SAG.

\begin{figure}[h]
    \centering
    \includegraphics[width=\linewidth]{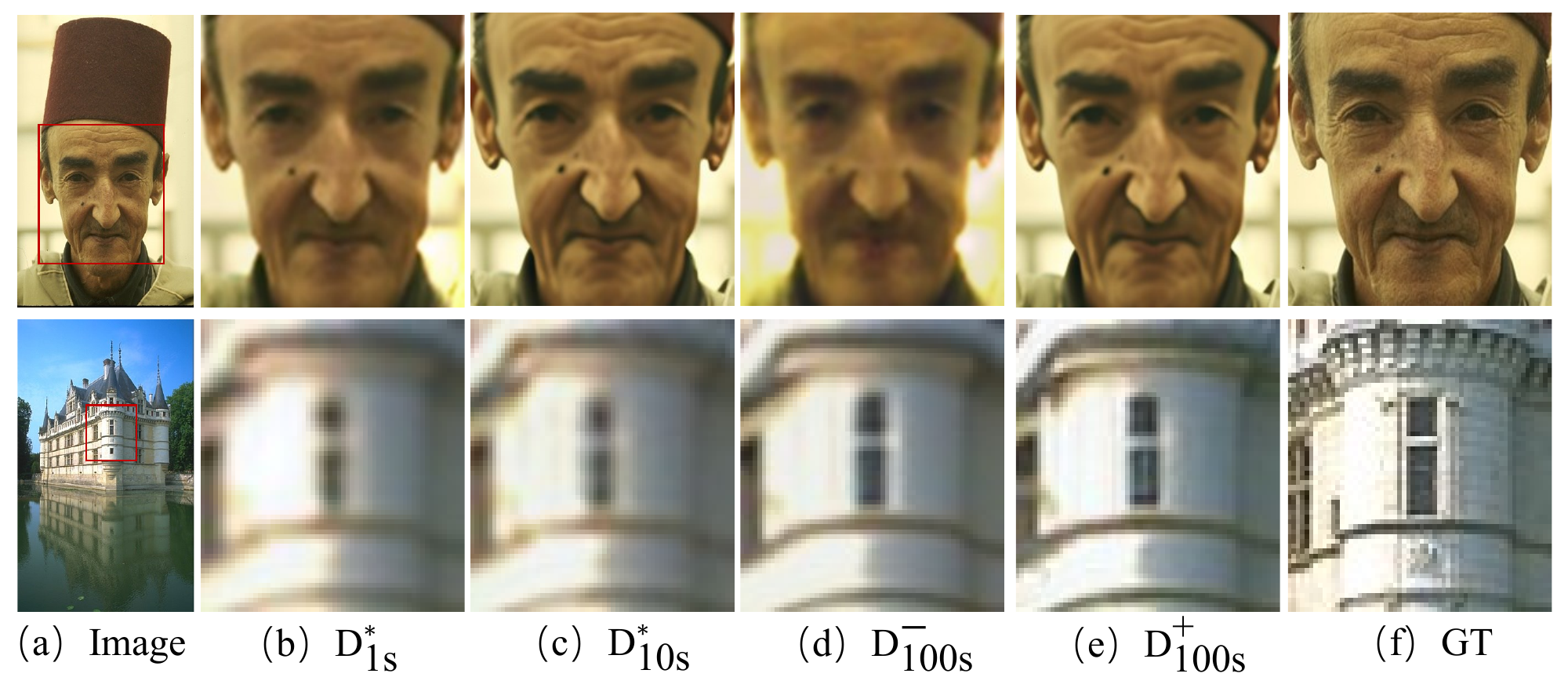}
    \vspace{-8.0mm}
    \caption{\textbf{Visual comparison of RGR denoising strategies impacting over-smoothing.} They are represented as `$\mathbf{D}_{1s}^{*}$' for continuous artifact refinement, `$\mathbf{D}_{10s}^{*}$' for refinement every ten steps, and `$\mathbf{D}_{100s}^{-}$' vs. `$\mathbf{D}_{100s}^{+}$' for pre- and post-100-step refinement.}
    \label{fig:guided_strategy_rgr}
    \vspace{-3.0mm}
\end{figure}

\begin{table}
    \centering
    \caption{\textbf{Comparison of different realistic latent udpate strategies.}  `$\mathbf{R}_{1s}^{*}$' refers to realistic-latent enhancement at each step and `$\mathbf{R}_{100s}^{+}$' indicates enhancement after 100 steps, respectively.}
    \vspace{-3.0mm}
    \label{tab:approach_to_update_latent_in_sargd}
    \begin{adjustbox}{max width=\linewidth}
    \begin{tabular}{c|l|ccc|ccc}
    \Xhline{1.5pt}
    \rowcolor{mygray}
    &
    &\multicolumn{3}{c|}{\textbf{Set14}}
    &\multicolumn{3}{c}{\textbf{B100}}\\
    \Xcline{3-5}{1.0pt} \Xcline{6-8}{1.0pt}
    \rowcolor{mygray}
     \multirow{-2}{*}{\textbf{Scale}}
    &\multirow{-2}{*}{\textbf{Type}}
    &\textbf{PSNR $\uparrow$}
    &\textbf{SSIM $\uparrow$}
    &\textbf{DISTS $\downarrow$}
    &\textbf{PSNR $\uparrow$}
    &\textbf{SSIM $\uparrow$}
    &\textbf{DISTS $\downarrow$}\\
    \hline \hline 
    \multirow{2}{*}{$\times 2$}
                    &$\mathbf{R}_{1s}^{*}$   &30.16 &0.821 &0.127 &30.29 &0.807 &0.138 \\
                    &$\mathbf{R}_{100s}^{+}$ &\textbf{31.40} &\textbf{0.842} &\textbf{0.126} &\textbf{30.92} &\textbf{0.817} &\textbf{0.132} \\

    \midrule
    \multirow{2}{*}{$\times 3$} 
                    &$\mathbf{R}_{1s}^{*}$   &30.33 &0.815 &0.129 &30.38 &0.792 &0.141 \\
                    &$\mathbf{R}_{100s}^{+}$ &\textbf{30.94} &\textbf{0.823} &\textbf{0.121} &\textbf{30.84} &\textbf{0.800} &\textbf{0.140} \\
    \midrule
    \multirow{2}{*}{$\times 4$} 
                    &$\mathbf{R}_{1s}^{*}$  &29.28 &0.770 &0.158 &29.92 &0.760 &0.158 \\
                    &$\mathbf{R}_{100s}^{+}$ &\textbf{30.01} &\textbf{0.778} &\textbf{0.156} &\textbf{30.23} &\textbf{0.763} &\textbf{0.156} \\
    \Xhline{1.5pt}
    \end{tabular}
    \end{adjustbox}
\vspace{-6.0mm}
\end{table}

\vspace{-4.0mm}
\paragraph{Effect of Varied Denoising Strategies within RGR.}
Table~\ref{tab:guided_strategy_rgr} demonstrates that the artifact refinement in the last 100 steps achieves the best results.
Analyzing the first two lines per scale shows that refining artifacts at each step improves outcomes over ten-step intervals, as seen with higher PSNRs (\textit{28.16 vs. 27.99 }on Set4 at $\times$2), due to variable artifact patterns in the latent space.
Comparing artifact refinement strategies indicates that post-100-step processing yields greater enhancements, as demonstrated by a PSNR increase from 26.94 to 28.33 at $\times$4 super-resolution.
Observations from panels (d) and (c) in Figure~\ref{fig:guided_strategy_rgr} suggest that artifact refinement early in the noisy latent phase causes more over-smoothing than later-stage refinement.

\vspace{-5.0mm}
\paragraph{Approaches to Updating the Realistic Latent in SAG.}
Table~\ref{tab:approach_to_update_latent_in_sargd} presents a comparative analysis of different strategies for updating the reference realistic-latent within SAG.
Notably, updating the realistic latent after 100 steps yields superior results across all metrics when compared to enhancements at every step. For instance, this strategy leads to an increase in PSNR from 30.16 to 31.40 at $\times$2, from 30.33 to 30.94 at $\times$3, and from 29.28 to 30.1 at $\times$4 on Set14. These findings suggest that enhancing the reference latent later in the process correlates with improved performance.

\vspace{-5.0mm}
\paragraph{Influence of Artifact Detection on SARGD.}
Table~\ref{tab:rgr_with_and_without_artifacts} contrasts SARGD's performance with and without the implementation of artifact detection in the latent space during inference, providing a line-by-line comparative analysis.
It is observed that the elimination of perceptual artifacts in the latent space significantly enhances super-resolution quality, as evidenced by improved metrics such as PSNR (\eg, \textit{30.84 vs. 30.06} on B100 at $\times$3 scaling) and DISTS (\eg, \textit{0.219 vs. 0.140} on B100 at $\times$3 scaling), highlighting the benefits of integrating artifact detection within SARGD.

\begin{figure}
    \centering
    \includegraphics[width=\linewidth]{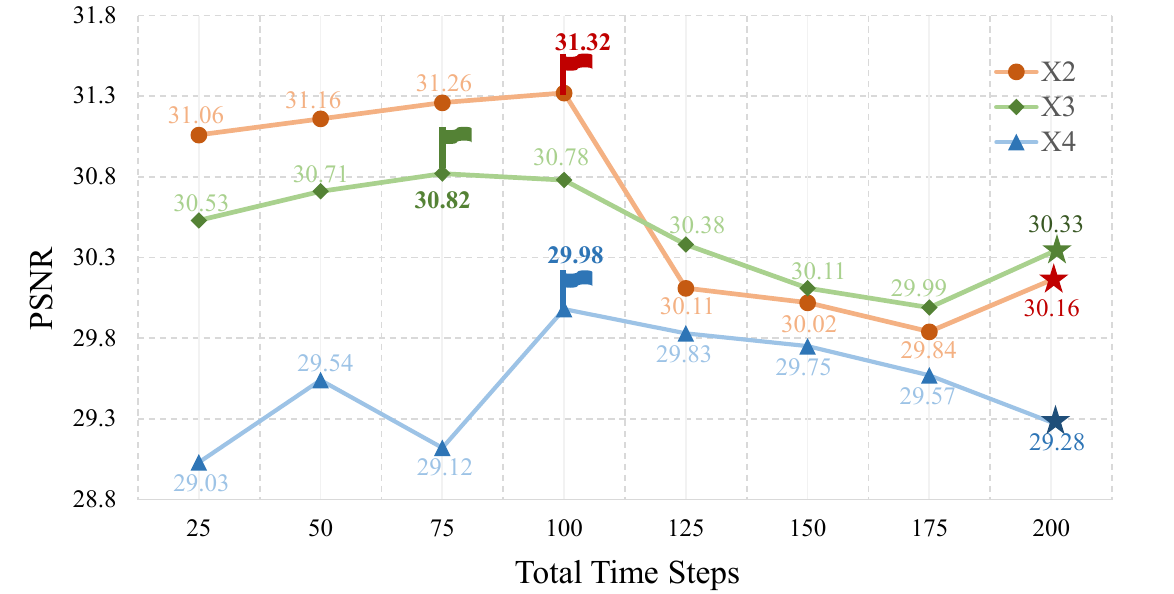}
    \vspace{-8.0mm}
    \caption{\textbf{Performance comparison of SARGD using different time steps} on Set14 for $\times$2, $\times$3, $\times$4 upscaling using PSNR. Our SARGD reaches peak score between 75 and 100 steps, effectively reducing the total inference steps by half, from 200 to 100.}
    \vspace{-5.0mm}
    \label{fig:inference_step_sargd}
\end{figure}

\vspace{-5.0mm}
\paragraph{Impact on Inference Steps for SARGD.}
Figure~\ref{fig:inference_step_sargd} illustrates that our proposed SARGD not only enhances image quality but also decreases the number of total inference steps required.
Specifically, the performance of SARGD in super-resolution steadily improves from the start and achieves optimal PSNR scores between 75 and 100 steps, such as a PSNR of 31.32 at 100 steps compared to 30.16 at 200 steps, which effectively cuts the inference time in half.
This suggests that the removal of artifacts during the diffusion phase serves to improve image quality in SR while simultaneously boosting the speed of the inference process.

\begin{table}
    \centering
    \caption{\textbf{Performance of SARGD with artifact detection (`$\mathcal{A}$') versus without} in latent space across $\times$2, $\times$3, and $\times$4 scales. Here, `\XSolidBrush' signifies the substitution of RGR with a direct sum of the latent and the initial realistic reference latent.}
    \vspace{-3.0mm}
    \label{tab:rgr_with_and_without_artifacts}
    \begin{adjustbox}{max width=\linewidth}
    \begin{tabular}{c|l|ccc|ccc}
    \Xhline{1.5pt}
    \rowcolor{mygray}
    &
    &\multicolumn{3}{c|}{\textbf{Set14}}
    &\multicolumn{3}{c}{\textbf{B100}}\\
    \Xcline{3-5}{1.0pt} \Xcline{6-8}{1.0pt}
    \rowcolor{mygray}
     \multirow{-2}{*}{\textbf{Scale}}
    &\multirow{-2}{*}{\textbf{$\mathcal{A}$}}
    &\textbf{PSNR $\uparrow$}
    &\textbf{SSIM $\uparrow$}
    &\textbf{DISTS $\downarrow$}
    &\textbf{PSNR $\uparrow$}
    &\textbf{SSIM $\uparrow$}
    &\textbf{DISTS $\downarrow$}\\
    \hline \hline 
    \multirow{2}{*}{$\times 2$} &\XSolidBrush    &31.24 &0.825 &0.138 &30.56 &0.801 &0.158 \\
                                &\Checkmark      &\textbf{31.40} &\textbf{0.842} &\textbf{0.126} &\textbf{30.92} &\textbf{0.817} &\textbf{0.132} \\

    \midrule
    \multirow{2}{*}{$\times 3$} &\XSolidBrush    &30.47 &0.796 &0.193 &30.06 &0.783 &0.219 \\
                                &\Checkmark      &\textbf{30.94} &\textbf{0.823} &\textbf{0.121} &\textbf{30.84} &\textbf{0.800} &\textbf{0.140} \\
    \midrule
    \multirow{2}{*}{$\times 4$} &\XSolidBrush    &29.98 &0.773 &0.234 &30.15 &0.753 &0.260 \\
                                &\Checkmark      &\textbf{30.01} &\textbf{0.778} &\textbf{0.156} &\textbf{30.23} &\textbf{0.763} &\textbf{0.156} \\
    \Xhline{1.5pt}
    \end{tabular}
    \end{adjustbox}
\vspace{-4.0mm}
\end{table}

\section{Conclusions}
\label{sec:conclusion}

In this study, we tackle the issue of artifacts that arise in diffusion-based super-resolution. Our SARGD incorporates two key mechanisms: Reality-Guided Refinement (RGR) and Self-Adaptive Guidance (SAG). RGR is designed to identify and correct artifacts within the latent space by utilizing a realistic latent as a reference, thereby improving the image's fidelity. 
To overcome the over-smoothing effect that comes from employing an initial realistic latent for guidance, SAG is introduced to refine the reference latent, improving the quality of the super-resolved images.

\noindent\textbf{Acknowledgements.} This work is supported by the National Natural Science Foundation of China (62271400), and the Shaanxi Provincial Key R\&D Program, China (2023-GHZD-02).

{
    \small
    \bibliographystyle{ieeenat_fullname}
    \bibliography{main}
}


\end{document}